\documentstyle[preprint,aps]{revtex}
\tightenlines
\begin{document}
\title{Ultra-refraction phenomena in Bragg mirrors}
\author{D. Felbacq, B. Guizal}
\address{LASMEA UMR 6602\\
Complexe des C\'{e}zeaux\\
63177 Aubi\`{e}re Cedex\\
France}
\author{F. Zolla}
\address{LOE ESA 6079\\
Facult\'{e} des Sciences de St-J\'{e}r\^{o}me\\
13397 Marseille Cedex 01\\
France}
\date{\today
}
\maketitle

\begin{abstract}
We show numerically for the first time that ultra-refractive phenomena do
exist in one-dimensional photonic crystals: we exhibit the main features of
ultra-refraction, that is the enlargement and the splitting of an incident
beam. We give a very simple explanation of these phenomena in terms of the
photonic band structure of these media.
\end{abstract}

It has recently been shown numerically as well as experimentally that near a
band edge, photonic crystals could behave as if they had an effective
permittivity close to zero \cite{dowl,enoch,bride}. Such a property induces
unexpected behaviors of light usually called ultra-refractive optics. The
main phenomena are the splitting or the enlargment of an incident beam, or a
negative Goos-H\"{a}nschen effect \cite{tayeb}. The common explanation of
these facts lie on the study of the photonic dispersion curves. Though
appealing, it seems difficult to turn this explanation into a rigorous one
as the notion of group velocity in a strongly scattering media seems
doubtful apart in the homogenization sense which is not the situation for
ultrarefractive optics. In our opinion, these surprising and beautiful
phenomena mainly rely on the rapid change in the behavior of the field
inside the structure when crossing a band edge. In this article, we provide
a rather simple explanation of some of these phenomena (splitting and
enlargment of an incident beam), which implies that they should be observed
with one dimensional structures (as foreseen by \cite{dowl}). Indeed, we
show by numerical experiments that it is the case in Bragg mirors (the
simplest photonic crystals).

From a theoretical point of view, we consider a periodic one dimensional
medium characterized by its relative permittivity $\varepsilon \left(
x\right) $, which is assumed to be real, illuminated by a plane wave. It is
well known that the band structure is determined by the monodromy matrix $%
{\bf T}$ of one layer \cite{lekner,oned}, that is, the matrix linking the
field and its derivative over one period. This matrix is a function of $%
\lambda $ and $\theta $. The main quantity is then $\phi \left( \lambda
,\theta \right) =\frac{1}{2}tr\left( {\bf T}\left( \lambda ,\theta \right)
\right) $. When $\left| \phi \left( \lambda ,\theta \right) \right| $ is
inferior to $1$ then $\left( \lambda ,\theta \right) $ belong to a
conduction band, and when $\left| \phi \left( \lambda ,\theta \right)
\right| $ is superior to $1$ then $\left( \lambda ,\theta \right) $ belong
to a gap. In fig. 1 we give a numerical example for a Bragg Mirror with $%
\varepsilon _{1}=1$, $\varepsilon _{2}=4$, $h_{1}=h_{2}=1$ (the lengths are
given in $\lambda $ units).

Now let us use a Gaussian beam as the incident field. Let us suppose that
the mean angle of the beam is zero (normal incidence) and that its
wavelength is very near a band edge. Then two things may happen. Reasoning
on the oriented wavelengths axis, if the beam is centered on the left side
of the gap (the dispersion diagram is given in the plane $\left( \lambda
,\theta \right) $, if one uses frequencies instead of wavelengths one has to
exchange left and right), the center of the beam belongs to a conduction
band and the edges of the beam belong to the gap. Consequently, after
propagation in the medium, the transmitted field has a narrowed spectral
profile, and therefore the beam is spatially enlarged (figures 1,2).
Conservely, if the beam is centered on the right side of the gap, then the
center of the beam belongs to the gap, and the edges of the beam belong to
the conduction band. Therefore, the transmitted field has two well separated
peaks and the beam is splitted in two parts (figures 1,3). The fundamental
remark here is that ultra-refractive phenomena are due to the rapid
variation of the conduction band with respect to the angle of incidence, 
{\it in complete contradiction with the habitual requested properties of
photonic crystals}, which are expected to have a dispersion diagram quite
independent of the angle of incidence.

Let us now check numerically the above explanations. We still use the
previous Bragg Mirror. The numerical experiments are done with an
s-polarized incident field of the form: 
\begin{equation}
u^{i}\left( x,y\right) =\int A\left( \alpha \right) \exp \left( i\alpha
x-i\beta \left( \alpha \right) y\right) d\alpha
\end{equation}
with $\alpha =k\sin \theta ,\alpha =k\sin \theta _{0},\beta \left( \alpha
\right) =\sqrt{k^{2}-\alpha ^{2}}$ and $k=2\pi /\lambda $, $\left| A\left(
\alpha \right) \right| =%
{\displaystyle{W \over 2\sqrt{\pi }}}%
\exp \left( -%
{\displaystyle{\left( \alpha -\alpha _{0}\right) ^{2}W^{2} \over 4}}%
\right) $. In all numerical experiments $W=0.5$, the variable $\theta _{0}$
is the mean angle of incidence.

In the first numerical experiment, we set $\lambda =2.7$ and $\theta _{0}=0%
{{}^\circ}%
$. We have plotted in figure (4a) the transmission coefficient as well as
the spectral profile of the transmitted beam. Obviously, this profile is
much narrower than the incident one. The map of the electric field is given
in figure (4b). The incident field is coming from below. As expected, we
observe a strong enlargement of the transmitted beam.

For the second numerical experiment, we use $\lambda =3$ and $\theta _{0}=0%
{{}^\circ}%
$. This time, the center of the beam belong to the gap. We have plotted in
figure (5a) the transmission coefficient as well as the spectral profile of
the incident and transmitted fields. It appears that there are two isolated
peaks, and therefore the transmitted field is splitted spatially into two
parts, as shown in figure (5b). At that point it is easily seen that by
switching the incident beam it is possible to keep only one transmitted
beam. This is done in the last experiment, where we set $\theta _{0}=10%
{{}^\circ}%
$ . As it can been seen on fig 6 (a), only the right part of the beam is
significantly transmitted, and thus there is only one transmitted beam (fig.
6 (b)). If Snell-Descartes law is directly applied to this situation, then
it seems that the medium has an optical index that is inferior to $1$.

As a conclusion, we have shown both theoretically and numerically that
ultra-refractive phenomena do happen in one-dimensional Bragg mirrors, or
more generally in one dimensional photonic crystals. They may be well
explained by means of the intersection of the support of the incident beam
with the gaps and the conduction bands. It must also be noted that, though
one dimensional photonic crystals exhibit ultra-refractive properties,
bidimensional or three dimensional ones should show a better efficiency due
their richer band diagrams. Nevertheless, doping 1-D structure or using
quasi-crystals may enable a fair control over the width of the gaps and
conduction bands, thus leading to the design of practical devices. Finally,
it should also be noted that such a surprising phenomenon as a negative
Goos-H\"{a}nchen effect does not seem to be possible in 1D structures.

\newpage

{\bf Figure captions:}

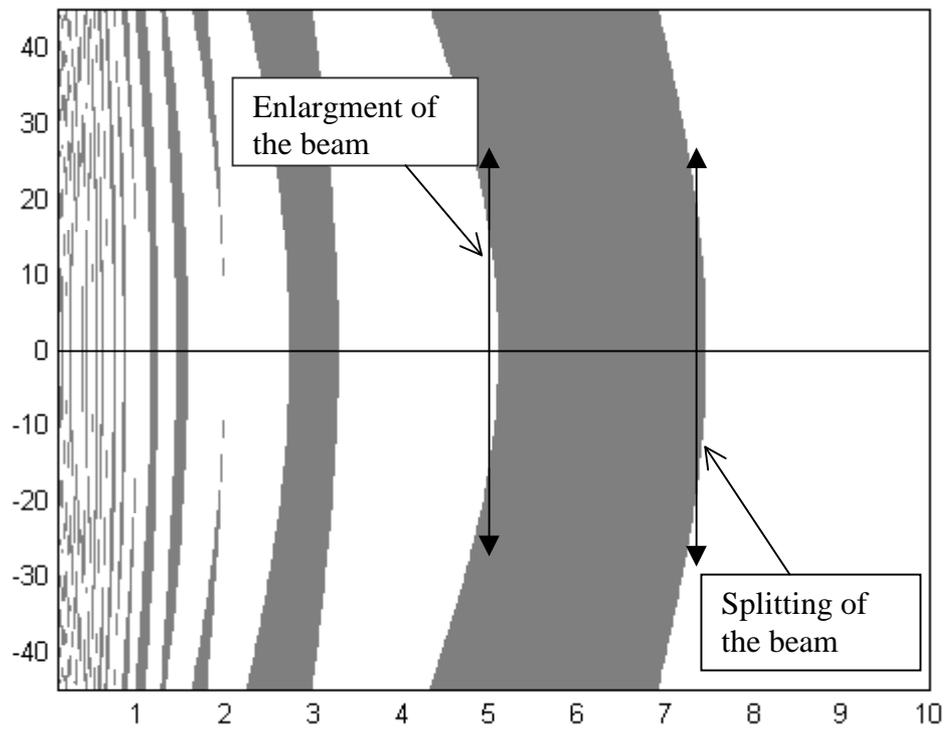
figure 1: Dispersion diagram of a Bragg mirror, with $\varepsilon
_{1}=1,\varepsilon _{2}=4,h_{1}=1,h_{2}=1$. The double arrowed lines
indicate the width of the Gaussian beams.

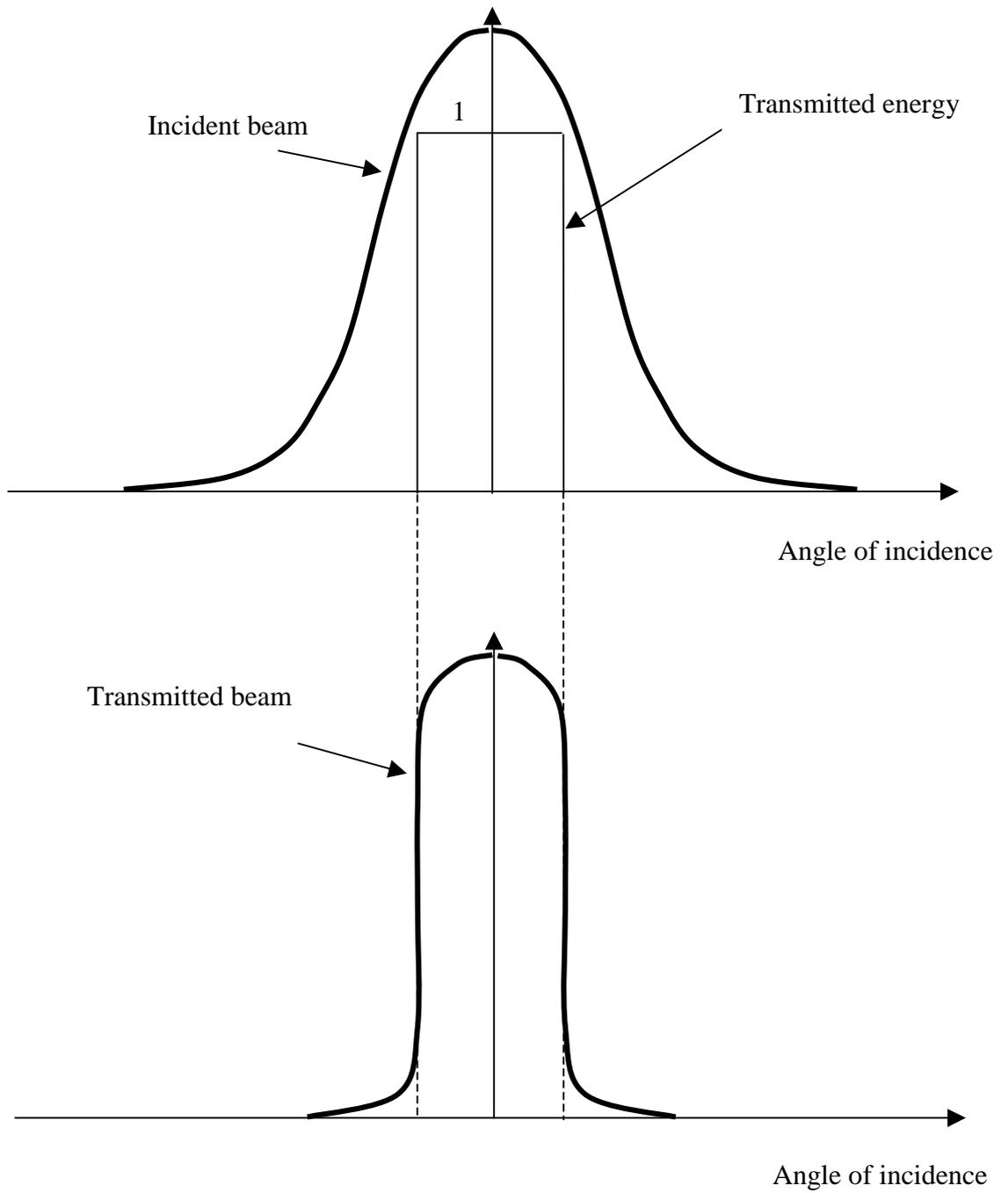
figure 2: Sketch of the behavior of the beam when spatially enlarged.

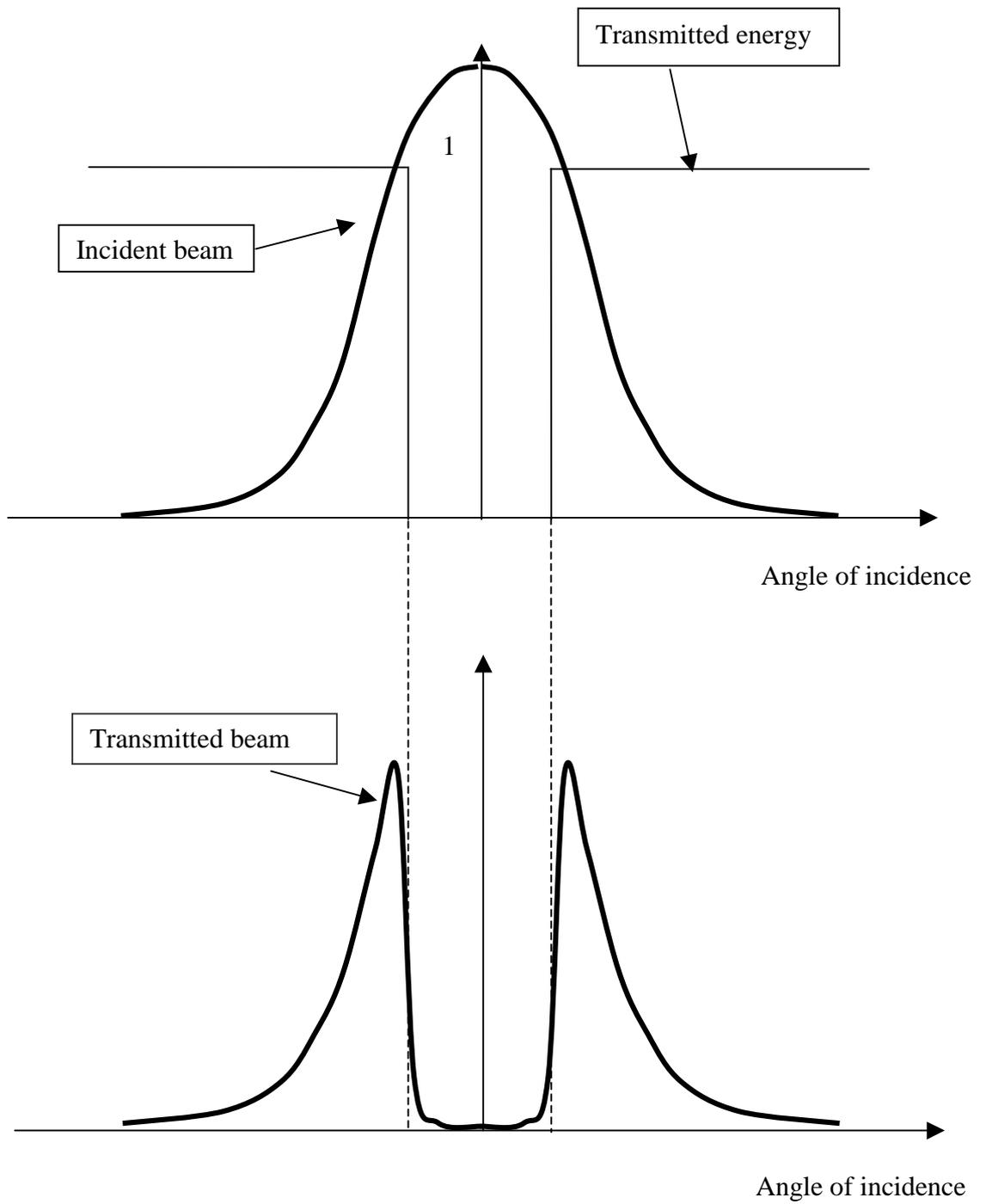
figure 3: Sketch of the behavior of the beam when splitted.

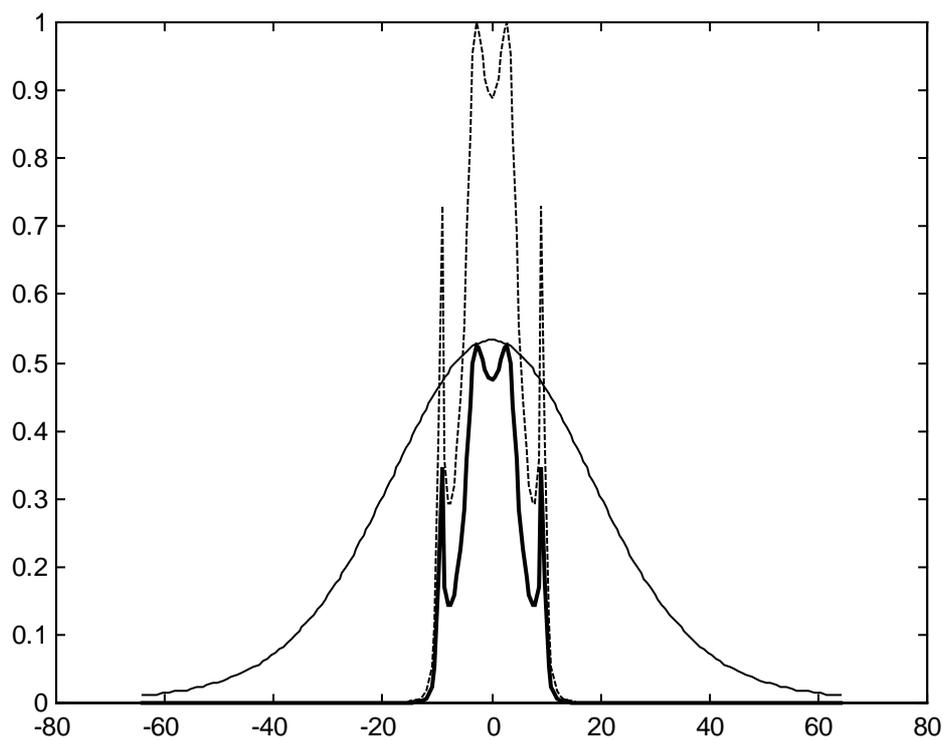
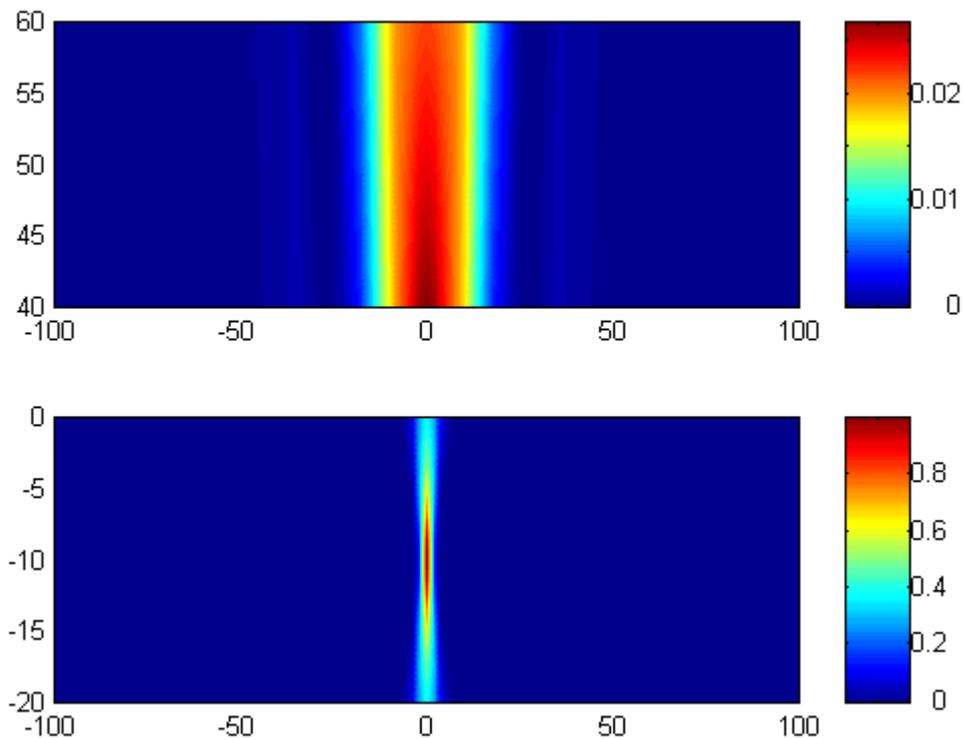
figure 4: (a) Transmission through the Bragg mirror vs. angle of incidence
(dotted line), spectral amplitude of the incident beam (solid line) and
spectral amplitude of the transmitted beam (thick line) ($\lambda
=2.7,\theta _{0}=0$).

\qquad \qquad (b) Map of the intensity of the electric field above and below
the Bragg mirror in the case of figure 2 (above: transmitted field, below:
incident field).

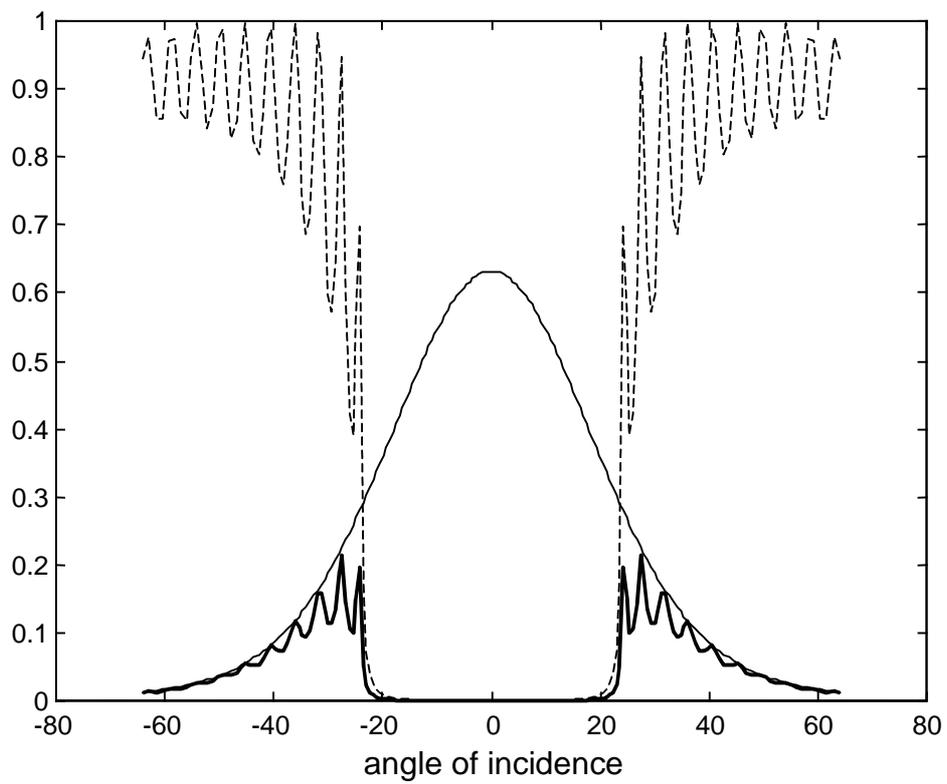
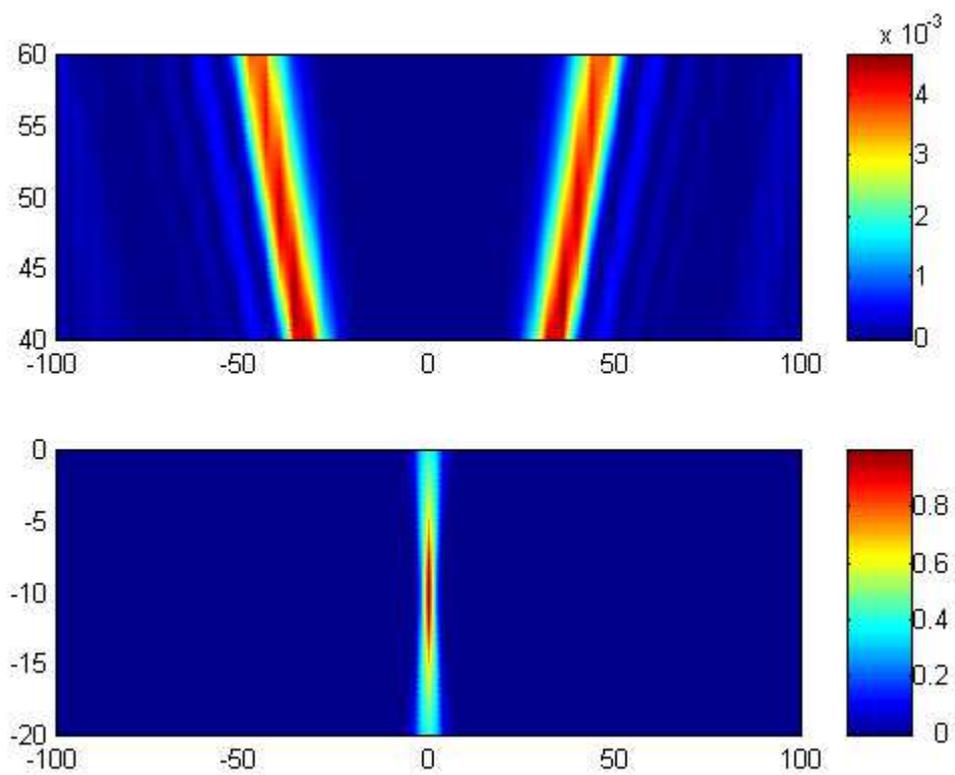
figure 5: (a) same as fig. 4 (a) in the case of figure 3 ($\lambda =3,\theta
_{0}=0%
{{}^\circ}%
$).

\qquad \qquad (b) Map of the intensity of the electric field above and below
the Bragg mirror in the case of figure 3 (above: transmitted field, below:
incident field).

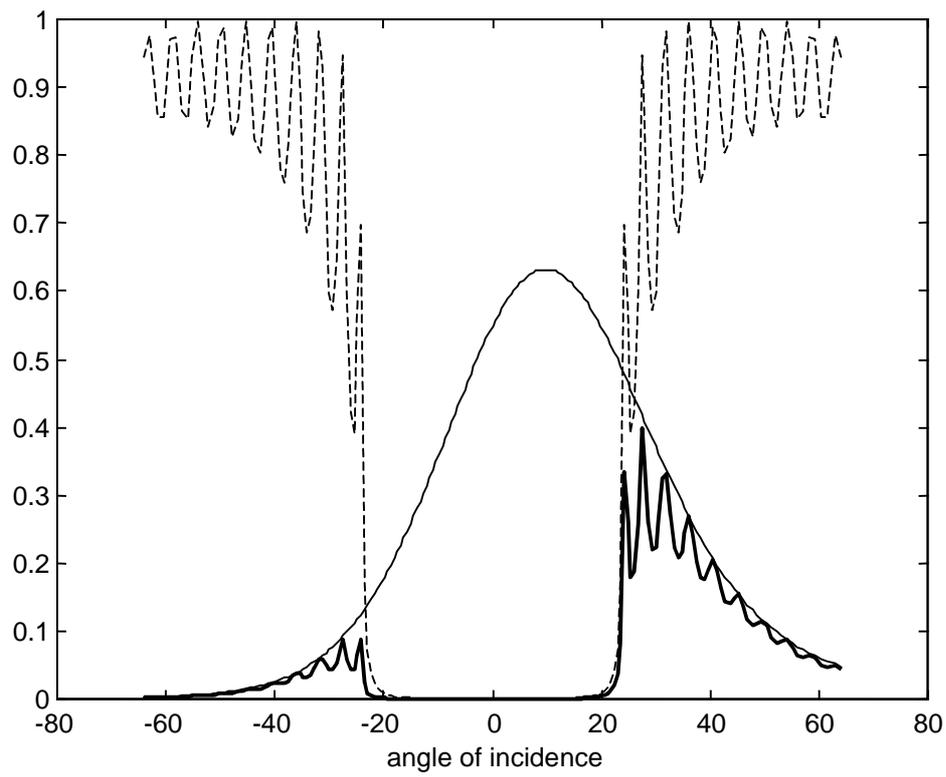
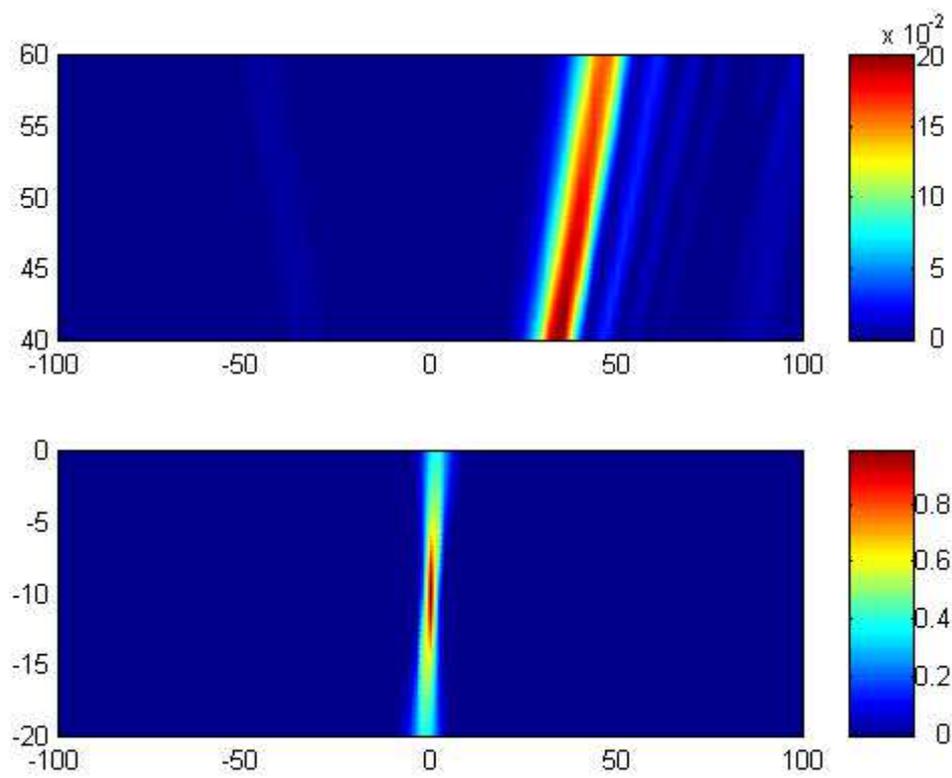
figure 6: (a) same as fig. 4 (a) in the case of figure 3.($\lambda =3,\theta
_{0}=10%
{{}^\circ}%
$).

\qquad \qquad (b) Map of the intensity of the electric field above and below
the Bragg mirror in the case of figure 3 (above: transmitted field, below:
incident field).

\end{document}